\documentclass[10pt,conference]{IEEEtran}

\usepackage{color}

\usepackage[pdftex]{graphicx}

\IEEEoverridecommandlockouts

\begin{document}

\title{Erasure Codes with a Banded Structure for Hybrid Iterative-ML Decoding}

\author{
\authorblockN{Alexandre Soro$^1$\thanks{This work was supported by the French ANR grant No 2006 TCOM 019
(CAPRI-FEC project).}, Mathieu Cunche$^2$, J\'er\^ome Lacan$^1$ and Vincent Roca$^2$}
\authorblockA{$^1$Univ. of Toulouse, ISAE/DMIA, 
10 avenue Edouard Belin,\\
BP 54032 - 31055 Toulouse cedex 4 - FRANCE\\
$^2$INRIA Rhone-Alpes, Plan\`ete team, Inovall\'ee, 655 av. de l'Europe, \\
Montbonnot, 38334 St Ismier cedex - FRANCE\\
Email: \{alexandre.soro,jerome.lacan\}@isae.fr, \{mathieu.cunche,vincent.roca\}@inria.fr}
}

\maketitle

\begin{abstract}
This paper presents new FEC codes for the erasure channel, LDPC-Band, that have been designed so as to optimize a hybrid iterative-Maximum Likelihood (ML) decoding.
Indeed, these codes feature simultaneously a sparse parity check matrix, which allows an efficient use of iterative LDPC decoding, and a generator matrix with a band structure, which allows fast ML decoding on the erasure channel.
The combination of these two decoding algorithms leads to erasure codes achieving a very good trade-off between complexity and erasure correction capability. 
\end{abstract}

\section{Introduction and Related Works}
\label{introduction}

For the transmission of data packets on erasure channels, linear binary FEC codes often offer the best compromise between fast encoding/decoding operations and a good level of erasure recovery capability. For example, random binary codes have a correction capability very close to channel capacity \cite{studholme:algorithmica}. Unfortunately, the decoding complexity of random codes is often prohibitive because the decoding algorithm, named Maximum-Likelihood (ML) decoding, basically consists in  solving a linear system, which can be done by inverting the associated matrix. This matrix represents the relations between the set of received symbols and the set of missing symbols, in our representation a row contains the coefficient used to build an encoding symbol. If $k$ denotes the dimension of the code the inversion has a complexity of $O(k^2)$ row operations. In the remaining of the paper the complexity will be evaluated in terms of row operation. A row operation includes the sum of two rows of the matrix and the sum of the corresponding symbols. It is importance to notice that in pratical applications the size of the symbols can be up to several hundred bytes. 

To reduce this complexity, Studholme and Blake \cite{studholme:isit06} showed that similar erasure capability can be obtained when the non zero entries of the generator matrix are located in a band of length $2\sqrt{k}$ and where each column contains $2\log(k)$ nonzero elements.  With this improvement, the complexity of the decoding is reduced to $O(k^{3/2})$ row operations.

The class of Fountain codes, like LT \cite{Luby2}  or Raptor \cite{raptor:Shokrollahi06} codes can also obtain very good  performance in terms of erasure correction capability. An estimation of the performance achieved by Raptor codes with ML decoding is provided in \cite{raptorCellular:2007}. However, despite the description of decoding algorithm provided in \cite{rfc5053:raptor}, the complexity needed to achieve this level of performance is not very clear.      

LDPC codes are another class of binary codes providing good level of decoding performance with extremely fast encoding/decoding algorithms \cite{ldpc:finiteLength}\cite{rfc5170}. Indeed, the classical iterative decoding algorithm for the erasure channel, based on the work of Zyablov \cite{ldpc:zyablov}, has a linear decoding complexity. The drawback of this algorithm is that it does not reach the performance of ML decoding. Moreover, the sparsity of the matrices also reduces the ML performance compared to full random matrices. 

Recently, two independent works  \cite{cunche08:hybrid}\cite{paolini08:CommLetters} proposed a hybrid iterative-ML decoding algorithm, where ML decoding is only used when the iterative decoding does not succeed to decode a received codeword.

The new codes we introduce in this paper also rely on a hybrid decoding.
Our goal is to build these codes in such a way that both the parity check matrix and the associated generator matrix have good properties for the iterative and ML decoding schemes.
More precisely, the parity check matrix should be sparse in order to yield good performance with the iterative decoding.
At the same time, the generator matrix should have a banded structure in order to reduce the computational complexity of the ML decoding scheme, as explained in \cite{studholme:isit06}. 
Thanks to a polynomial representation of the rows and the columns of the generator and parity check matrices, we introduce a method that enabled us to build such codes.

The idea of window based encoding in LDPC codes has been proposed by Haken, Luby and al.
in patent \cite{USPatent6486803}.
However, the goal of this patent is only to minimize memory access during encoding, by
localizing the access in a window that slides over the input file.
Independantly of whether our proposal falls into the scope of this patent or not,
we see that, from a purely scientific point of view, the goal of \cite{USPatent6486803}
completely departs from the approach discussed in the current paper, as well as the
theoretic tools we introduce to achieve our goals.

The paper is organized as follows:
we detail the polynomial approach for the construction of our LDPC-Band codes in Section~\ref{sec:ConstructionCode}.
In Section~\ref{sec:correctionCapabilityAnalysis}, we analyse the erasure correction capability of the proposed code and we provide simulation results.
Finally, in Section \ref{sec:complexityAnalysis} we evaluate the complexity of the hybrid decoding. Then we conclude.

\section{Construction of the code}
\label{sec:ConstructionCode}
\subsection{General structure of the code}
\label{sec:General_structure}

Let $\mathcal{G}$ be the systematic binary generator matrix of the banded code:
\begin{displaymath}
\mathcal{G}=(Id|M)
\end{displaymath}
Let $\mathcal{H}$ be its associated parity check matrix:
\begin{displaymath}
\mathcal{H}=(A|U)
\end{displaymath}
Let $k$ be the dimension code and $n$ the length of the code. It follows that $\mathcal{G}$ is a $k\times n$ matrix, $A$ is a $(n-k)\times k$ matrix and $U$ is a $(n-k) \times (n-k)$ binary matrix. 


Let us define $U$ as a lower triangular Toeplitz matrix defined by its first column $(1, u_1,\ldots,u_{n-k-1})$. Since all the diagonal elements are equal to $1$, $U$ is a full rank matrix. In addition, let us consider the associated polynomial  $u(x) = 1+\sum_{k=1}^{n-k-1}{u_{k}x^{k}}$. 
The coefficient of $U$ are the following:

\begin{displaymath}
\{u_{i,j}\} =\left\{ 
\begin{array}{ccc}
u_{i-j} & if & i > j \\ 
0 & if & Ã— i < j \\
1 & if & i = j
\end{array}
\right .
\end{displaymath}

\begin{center}
$U =
\left( \begin{array}{cccccccc}
  1 & 0 & 0 &  & \ldots &  & 0 & 0\\
  u_1 & 1 & 0 &  &  &  &  & \\
  u_2 & u_1 & 1 &  &  &  & \vdots & \vdots\\
  \vdots_{} & u_2 & u_1 &  &\ddots  &  &  & \\
  u_i & \vdots_{} & u_2 &  &  &\ddots  & 0 & \\
  \vdots & u_i & \vdots_{} &  &  &  & 1 & 0 \\
  u_{n-k-1} & \vdots & u_i &  &  &  & u_1 & 1 \\
\end{array}\right)$
\end{center}

The relations between the generator matrix $\mathcal{G}$ and its associated parity check matrix $\mathcal{H}$ : $\mathcal{G} \times \mathcal{H}^T  = 0$ give:
$$
M = {(U^{-1}A)}^{T}$$
$$
UM^{T} = A
$$


Let $B$ be the band width of the band matrix $M$ studied here, which will have the following form:

\begin{center}

$M = 
\left( \begin{array}{ccccccc}
  m_{0,0} & m_{0,1} & \ldots & m_{0,B-1} & 0 & \ldots & 0\\
  0 & m_{1,0} & m_{1,1} & \ldots & m_{1,B-1} & 0 & \ldots\\
  \vdots & 0 & \ddots & \ddots & \ldots & \ddots & 0\\
\end{array}\right)
$
\end{center}

For $i=0,\ldots,k-1$, let us define the polynomial  $m_i(x)=\sum_{j=0}^{B-1}{m_{i,j}x^{j}}$ corresponding  to the $i^{\textrm{th}}$ row of $M$. 
Since $A=UM^{T}$, it can be shown that, for $i=0,\ldots,k-1$, the  $i^{\textrm{th}}$ column of $A$ is characterized by the polynomial $a_i(x)$ such that $a_i(x)=u(x)m_i(x)$ (see Figure \ref{fig:parityCheckMatrix}).
This result is important as it allows to construct banded generator matrices from particular distributions in the parity check matrix.
Let us define the Hamming weight of a polynomial as the number of monomials.
Then, the number of non-zero elements of a column of $A$ is the weight of $a_i(x)$, noted $W(a_i)$. 

\subsection{Design of the matrices}
\label{sec:DesignMatrices}

In order to build a code supporting hybrid iterative/ML decoding, a first constraint is to optimize the iterative decoding on the parity check matrix.
This decoding is very sensitive upon the degree of the symbol and check nodes, i.e. the weight of the columns and the rows of the parity check matrix.
Thanks to the polynomial approach, these parameters can be controlled. Indeed the weight of the columns and rows of $U$ is defined by the number of monomials in $u(x)$, and the weight of the column of $A$ is defined by the number of monomials in the different $a_i(x)$. 
We will see later how the row weight can be controlled.
The second constraint concerns the ML decoding. As explained before, the generator matrix of this code must be band of width $B$ in order to reduce the ML decoding complexity.
We do not impose any other constraint on the generator matrix as we assume that the matrix is sufficiently random to allow good correction capabilities under ML decoding.

In order to support an efficient iterative decoding, the columns of the parity check matrix must have a small weight.
In particular we must choose a small weight polynomial for $u(x)$, the polynomial defining $U$ the right side of the parity check matrix. 

The main point of the process is to find polynomials ${m_i(x)}$ of degree close or equal to $B$, such that $a_i(x)$ have few monomials, where $a_i(x)=m_i(x)u(x)$.
Let us call these polynomials \emph{candidate polynomials}.
Finding these candidate polynomials can be achieved with an exhaustive search, in advance, once the various code parameters (n, k, B) are known. 
The details of how this search is done does not impact the decoding efficiency and therefore are not described in the reminder of this paper.

\begin{figure}

$\mathcal{H} = \left( \begin{array}{ccccc|cccc}
  a_{0, 0} &  &  &  &  & u_0 &  &  & \\
  a_{0, 1} & a_{1, 0} &  &  &  & u_1 & u_0 &  & \\
  a_{0, 2} & a_{1, 1} & a_{2, 0} &  &  & u_2 & u_1 & u_0 & \\
  a_{0, 3} & a_{1, 2} & a_{2, 1} &  &  & u_3 & u_2 & u_1 & \\
  \vdots & a_{1, 3} & a_{2, 2} & \ddots &  & \vdots & u_3 & u_2 & \ddots\\
  & \vdots & a_{2, 3} &  &  &  & \vdots & u_3 & \\
  &  & \vdots &  &  &  &  & \vdots &  \\
  &  &  &  &  &  &  &  & 
\end{array} \right)$\caption{The parity check matrix of the code with a band structure}
\label{fig:parityCheckMatrix}
\end{figure} 

Once a polynomial $u(x)$ and a set of candidate polynomials have been selected, the weight of the columns of parity check matrix is fixed. But the weight of the rows can still be modified to fit a specific value. Indeed by permuting two polynomial in $M$, or by exchanging a polynomial with another candidate polynomial of same weight, we can modify the weight of several rows without changing the weight of the columns.

\subsection{Optimizations}
\label{sec:Optimisations}

We have now defined a way to implement banded codes for both decoding techniques. However this implementation is limited to a number of encoding symbols of exactly $n=2k+B$. Furthermore, these codes turn out to be quite inefficient for the ML decoding in terms of erasure recovery capability, because the first non-systematic symbols only protect a small part of the source symbols. 
This is a negative side effect of the approach. We now explain how to solve this problem.

Given a polynomial $u(x)$ and a bandwidth $B$, the set of candidate polynomials usualy contains several elements.
Therefore it is possible to choose different polynomials and to interlace them to build the matrix $M$, and the matrix $A$ inherits from a corresponding interlacement in its polynomial.
This technique has two main benefits.
The first one is an improvement of the decoding performances for the ML decoding, with limited impacts on the iterative decoding.
Indeed, many polynomials interlaced produce more variety in the generator matrix and avoid diagonals and regularities that damage the performances of the ML decoding.
The second benefit is the suppression of the side effect mentionned above.
The idea is to suppress the $\frac{B}{2}$ first and last columns of $M$.
As it breaks the polynomial relations between $A$, $U$ and $M$ because of the first and last $\frac{B}{2}$ lines of $M$ that are now truncated, we replace these lines with candidate polynomials which are of degree $\frac{B}{2}$ at maximum, in order to keep the matrix banded, and keep the constraints upon $A$ and $U$.
As a consequence, the encoding packets at the edge of the non-systematic part of the generator matrix are now connected to more source symbols. This also suppress the influence of the band size on the code rate, as the number of symbols produced is now $n=2k$, meaning a code rate strictly equal to $1/2$.

\begin{figure}[htb]
  \begin{center}
   \includegraphics[width=.2\textwidth]{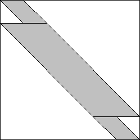}
    \end{center}
    \caption{Design of the banded matrix of the non-systematic part with side effect removed}
   \label{fig:trunc}
\end{figure}

The last point is to handle different code rates. To that goal, we define the $f$ parameter which corresponds to the column shift between consecutive lines of $M$. For instance, for a ratio of $r=1/2$, each successive line is shifted by $1$ position from the previous one, i.e. $f=1$. In order to handle other ratios, we need to take $f=1/r-1$. Moreover, in this case, the first and last $\frac{B}{2f}$ lines are of degree $\frac{B}{2}$.
If $f$ is not an integer, we should use a family of integer offsets $\{f_i\}$ such as the average value of this set is equal to $f$.

To summarize, in this section we have defined a way to create erasure band codes that handle efficiently (both in terms of erasure recovery capability and algorithmic complexity) both iterative (thanks to the sparseness of the parity check matrix) and ML decoding (thanks to its banded structure), for different code rates.

\section{Correction Capability Analysis}
\label{sec:correctionCapabilityAnalysis}
\subsection{Theoretical Analysis}

\paragraph*{Influence of the band on the ML performances}
For the ML performances on the generator matrix $\mathcal{G}$, the crucial point is to determine a band size that ensures both a quick decoding and good correction capabilities. With the first constraint, the band size used should be as narrow as possible. However, this size should also be sufficient in order to avoid an explosion of the decoding overhead. It is important to notice that the band size corresponds to the maximum degree of the associated binary polynomials ${m_i(x)}$.

Studholme and Blake \cite{studholme:isit06}\cite{studholme:algorithmica} have studied the influence of the window size on binary random matrices constrainted in a window. The conclusion of their work is that for random binary matrices constrained in a window, its size has to be kept close to $2\sqrt{k}$. Moreover, in this window, the number of non-zero elements must be at least $2\log{k}$. These values guarantees that the probability of being full rank is close to the one of pure random matrices. As our approach uses specific configurations of banded matrices, this value should be considered as a minimum. The band size in our case has to be kept as low as possible and for long dimensions, as close as possible to this value, because it will have a great impact on decoding speed. The size of the band will then be a tradeoff between a good ML decoding speed and a strong ML decoding performance.

\paragraph*{Density evolution analysis}

A density evolution approach \cite{Rich-Shok-Urba} for the design of this code has been studied, but have faced many obstacles. 
If we choose to consider a infinite matrix with an infinite band width, the density evolution will give the same
distributions as for a standard Repeat-Accumulate code. But building a finite length matrix having a such distribution can be difficult
as the rows distribution strongly depends on the choice of the polynomials and their permutations. Furthermore as the band is small, a
strong irregularity in the joint degree distribution  \cite{Kasai2003} may appear. 

On the other hand if we choose an infinite matrix with a finite band width, the hypothesis of infinite length cycle 
does not hold anymore.

\subsection{Simulation Results}

In this section, we compare the correction capability of our codes with standard LDPC-Staircase codes. First of all, it is worth noticing that we will use band codes with multiple polynomials interlaced. Indeed, experiments have illustrated the fact that using only one polynomial degrades greatly the decoding performances and will not be discussed here.

The crucial point is now to select a set of polynomials that will respect the conditions in the band, but also that have great performances in both decoding. In this study, we have tried to keep as close as possible of iterative adapted distributions. However, because of the additional conditions of banded matrices, an evolution density approach is not straightforward. It is also worth noticing that the chosen set of polynomials is always a compromise between performance of iterative decoding and ML decoding. This allows the user to adapt the polynomials to the environment of the channel. Experiments have also shown that the way the polynomials are interlaced and their position has only a little impact on the ML decoding performances, even for a random interlacing. The results presented in Table \ref{tab:ineff} and Table \ref{tab:ineff2} are based on a regular interlacing for $B=100$ and $B=200$ for a code rate of $1/2$. We study here, the average overhead required in order to successfully decode the received symbols.

\begin{table}[ht]
\begin{center}
\footnotesize{
\begin{tabular}{|r|c|c|c|c|}
\hline
\emph{Decoding scheme} & Iterative &  Maximum-Likelihood \\
\hline
LDPC-Staircase $N_1=5$	& 14.24\% &  1.21\% \\
\hline
LDPC-Band - B = 100     & 18.39\% &  2.97\% \\
\hline
LDPC-Band - B = 200     & 14.75\% &  1.24\% \\
\hline
\end{tabular}}
\end{center}
\caption{Average overhead W.R.T decoding scheme, k=1000 code rate=1/2 }
\label{tab:ineff}
\end{table}

\begin{table}[ht]
\begin{center}
\footnotesize{
\begin{tabular}{|r|c|c|c|c|}
\hline
\emph{Decoding scheme} & Iterative &  Maximum-Likelihood \\
\hline
LDPC-Staircase	$N_1=5$ & 13.95\% &  1.15\% \\
\hline
LDPC-Band - B = 200     & 16.23\% &  1.19\% \\
\hline
\end{tabular}}
\end{center}
\caption{Average overhead W.R.T decoding scheme, k=2000 code rate=1/2 }
\label{tab:ineff2}
\end{table}

As expected, the band size is a crucial factor for the decoding performances. A band size of $100$ is less performant but will have a greater decoding speed. In order to have efficient systems, we can see that the band size has to be large compared to the Windowed Erasure Codes. These results also show that LDPC-Band codes are really close to standard LDPC-Staircase codes, while being more constrained.

\section{Complexity Analysis}
\label{sec:complexityAnalysis}
\subsection{Theoretical Analysis}

A key point in data transmission is the encoding and decoding speed of the code.
In this study we do not include the creation of the code in the complexity analysis.
Indeed, the generation of the generator and parity check matrices is straightforward and 
can be done out of line.
On the decoder side, both decoding algorithms apply.
The first one is an iterative decoding on the parity check matrix. This decoding is fast and has a linear complexity on the dimension of the code $O(k)$.
When the iterative decoding fails, a ML decoding is used on the generator matrix.
The complexity of this algorithm is in the general case $O(k^2)$ row operations.
However the decoding of LDPC-Band codes benefits from the band structure.
Thanks to this structure, an optimized LU decomposition of the matrix \cite{Golub:1996} 
leads to a complexity of $O(kB)$ row operations.
This means that for $k=2000$ source symbols and a band size of $B=200$ symbols, the theoretical speed gain compared to a classical ML decoding will be about one order of magnitude. The complexity obtained here has to be compared with the complexity of $O(k^{3/2})$ row operations of the Studholme and Blake approach.

However, in practice, the size of the matrices that need to be inverted is lower than $k$.
There are three reasons:
(1) whenever a source symbol is received, the corresponding row in the generator matrix is removed.
Then (2), only the columns of the non-systematic part that are received are used.
It means that for a code rate $r$ the system that has to be inverted is only in the
order of $(1-r)k \times (1-r)k$. 
Finally (3) the iterative decoding may have rebuilt some missing symbols, thereby
reducing the system size.


\subsection{Simulation Results}

We carried out several tests to assess the computation benefits of our proposal.
These tests were obtained on a 4xIntel Xeon5120 @ 1.86GHz/4 GB RAM/Linux PC.
We compared the LDPC-Band codes with a band width of 200 with two other codes: 
\begin{enumerate}
  \item  the LDPC-Staircase codes, regular repeat accumulate codes standardized by the IETF
	\cite{rfc5170}.
	Following the optimisation for the hybrid iterative/ML decoding proposed in
	\cite{cunche08:gaussian}, the degree of the source symbol nodes is set to 5. 
  \item  the random windowed codes proposed in \cite{studholme:isit06}.
	These codes can be seen as a non systematic LDGM codes with $2log(k)$
	elements per column.
	Because it is a non-systematic code, iterative decoding cannot be used (unlike standard LDGM codes).
 \end{enumerate}

\begin{figure}
\centering
\includegraphics[width=3.6in]{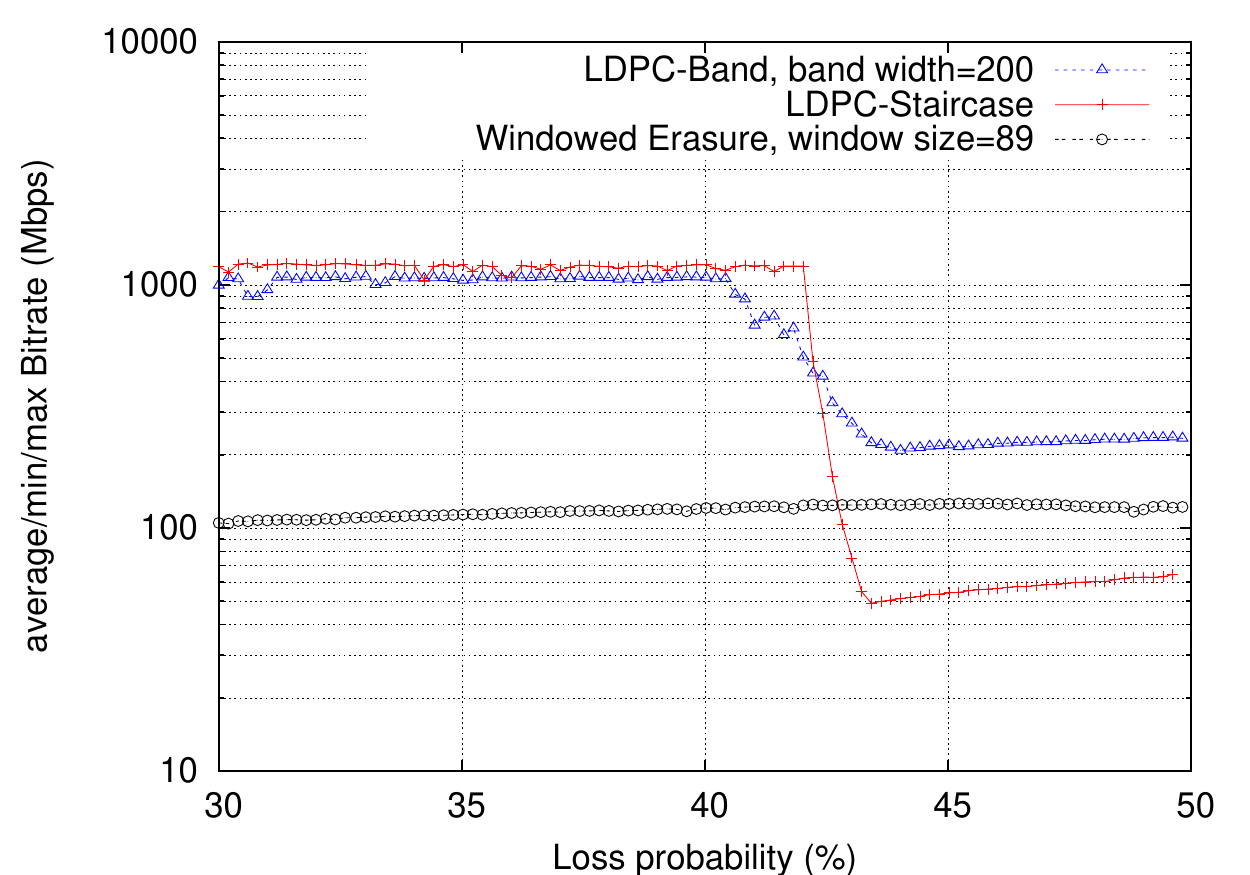}
\caption{Bitrate W.R.T loss probability, k=2000 code rate=1/2 and symbol size=1024 bytes}
\label{fig_bitrate_loss_prob}
\end{figure}
Figure \ref{fig_bitrate_loss_prob} shows the average decoding bitrate
as a function of the loss probability of the channel for the various code. 
The LDPC-Band and the LDPC-Staircase performances can be divided into two parts.
In the region where the loss probability is low, the iterative decoder is sufficient to recover the losses and the bitrate is high.
In the region of high loss probability the ML decoder is needed and the bitrate
is reduced.
On the opposite, the decoding speed of Windowed code does not depend on the loss
probability, since it can only be decoded with a ML decoder. 

We see that when the iterative decoding is used, the LDPC-Staircase and the LDPC-Band
present comparable decoding speed (with a slight advantage to LDPC-Staircase codes),
and they are both significantly faster than Windowed codes (that need ML decoding). 

When ML decoding is needed, LDPC-Band codes outperform the other two codes:
they are roughly 2 times faster than the Windowed codes, and 
roughly 4 times faster than LDPC-Staircase codes.
Let us remind that the ML decoding of the LDPC-Band and the LDPC-Staircase is equivalent to solve a linear system of size $(1-r)k \times (1-r)k$, whereas 
the ML decoding of the windowed erasure codes requires to solve a system of size $k \times k$
(being non systematic, no source symbols is received and the system still has $k$ unknown
variables).

Table \ref{tab:ML_decoding_time} present the average ML decoding time for the three compared codes
as a function of the number of source symbols $k$. The LDPC-Band is always faster than the two other 
codes. Furthermore the ML decoding speed falls more quickly with the two other codes than with the LDPC-Band.
The decoding speed with an iterative decoder are shown in Table \ref{tab:IT_decoding_time}. The LDPC-Band
is slower than the LDPC-Staircase but remains very close to it under iterative decoding.

\begin{table}[ht]
\begin{center}
\footnotesize{
\begin{tabular}{|r|c|c|c|}
\hline
\emph{k}  &  1,000 &  2,000 &  4,000    \\
\hline
LDPC-Band	 &  326 Mbps &  235 Mbps  &   150 Mbps \\
\hline
LDPC-Staircase      &  125 Mbps &   60 Mbps &   30 Mbps \\
\hline
Windowed Erasure  &  220 Mbps &  120 Mbps  &   68 Mbps\\
\hline
\end{tabular}}
\end{center}
\caption{Average ML decoding time as a function of the object size, for a code rate=1/2 and symbol size = 1024 bytes.}
\label{tab:ML_decoding_time}
\end{table}

\begin{table}[ht]
\begin{center}
\footnotesize{
\begin{tabular}{|r|c|c|c|}
\hline
\emph{k} &  1,000 &  2,000 &  4,000    \\
\hline
LDPC-Band	&  1100 Mbps &  1050 Mbps  &   900 Mbps\\
\hline
LDPC-Staircase       &  1300 Mbps &   1200 Mbps &   1000 Mbps \\
\hline
\end{tabular}}
\end{center}
\caption{Average iterative decoding time as a function of the object size, for a code rate=1/2 and symbol size = 1024 bytes.}
\label{tab:IT_decoding_time}
\end{table}

\section{Conclusion}
In this paper we proposed a flexible scheme that allows the construction of generator and parity check matrices and their efficient hybrid decoding.
The iterative decoding on the parity check matrix provides a fast way to recover the source symbols, as decoding complexity is linear.
When the environment is too harsh and the loss rate is close to the recovery capability of the codes, a Maximum-Likelihood decoding is applied on the structured generator matrix.
Thanks to the band structure of this matrix, the ML decoding complexity is reduced to $O(kB)$ row operations.
This property allows the hybrid decoder to sustain high decoding speeds, even in high loss environments.

Thanks to the polynomial representation of the matrices, we presented a practical way of building the matrix of such codes.
Our results shows that our LDPC-band codes match the performances of LDPC-staircase codes in terms of iterative decoding speed,
while being much faster than these codes when ML decoding is required.
Furthermore the LDPC-band codes exhibit erasure recovery capabilities close to standard repeat-accumulate codes using iterative
decoding, and are close to the channel capacity when decoded with a ML scheme.
In future works, a density evolution study should enable us to further improve the erasure correction  capabilites of the iterative decoding scheme, without degrading the ML decoding capability. 



%

\bibliographystyle{IEEEtran}
\bibliography{bandMatrices}

\begin{thebibliography}{10}
\providecommand{\url}[1]{#1}
\csname url@rmstyle\endcsname
\providecommand{\newblock}{\relax}
\providecommand{\bibinfo}[2]{#2}
\providecommand\BIBentrySTDinterwordspacing{\spaceskip=0pt\relax}
\providecommand\BIBentryALTinterwordstretchfactor{4}
\providecommand\BIBentryALTinterwordspacing{\spaceskip=\fontdimen2\font plus
\BIBentryALTinterwordstretchfactor\fontdimen3\font minus
  \fontdimen4\font\relax}
\providecommand\BIBforeignlanguage[2]{{%
\expandafter\ifx\csname l@#1\endcsname\relax
\typeout{** WARNING: IEEEtran.bst: No hyphenation pattern has been}%
\typeout{** loaded for the language `#1'. Using the pattern for}%
\typeout{** the default language instead.}%
\else
\language=\csname l@#1\endcsname
\fi
#2}}

\bibitem{studholme:algorithmica}
C.~Studholme and I.~Blake, ``Random matrices and codes for the erasure
  channel,'' \emph{Algorithmica}, April 2008.

\bibitem{studholme:isit06}
------, ``Windowed erasure codes,'' \emph{Information Theory, 2006 IEEE
  International Symposium on}, pp. 509--513, July 2006.

\bibitem{Luby2}
M.~Luby, ``${LT}$ codes,'' in \emph{Proceedings of the 43rd Annual IEEE
  Symposium on Foundations of Computer Science, FOCS'2002}, 2002, pp. 271--280.

\bibitem{raptor:Shokrollahi06}
A.~Shokrollahi, ``Raptor codes,'' \emph{IEEE Transactions on Information
  Theory}, vol.~52, no.~6, pp. 2551--2567, 2006.

\bibitem{raptorCellular:2007}
M.~Luby, T.~Gasiba, T.~Stockhammer, and M.~Watson, ``Reliable multimedia
  download delivery in cellular broadcast networks,'' \emph{Broadcasting, IEEE
  Transactions on}, vol.~53, no.~1, pp. 235--246, March 2007.

\bibitem{rfc5053:raptor}
M.~Luby, A.~Shokrollahi, M.~Watson, and T.~Stockhammer, ``{Raptor Forward Error
  Correction Scheme for Object Delivery},'' Internet Engineering Task Force
  (IETF), Tech. Rep. RFC 5053 (Proposed Standard), Oct. 2007.

\bibitem{ldpc:finiteLength}
C.~Di, D.~Proietti, T.~Richardson, E.~Telatar, and R.~Urbanke, ``Finite length
  analysis of low-density parity-check codes on the binary erasure channel,''
  \emph{{IEEE} {T}ransactions on {I}nformation {T}heory}, vol.~48, pp.
  1570--1579, 2002.

\bibitem{rfc5170}
V.~Roca, C.~Neumann, and D.~Furodet, ``{{Low Density Parity Check (LDPC)
  Staircase and Triangle Forward Error Correction (FEC) Schemes}},'' Internet
  Engineering Task Force (IETF), RFC RFC 5170 (Proposed Standard), june 2008.

\bibitem{ldpc:zyablov}
V.~V. Zyablov and M.~S. Pinsker, ``Decoding complexity of low-density codes for
  transmission in a channel with erasures,'' \emph{{Probl. Peredachi Inf.}},
  vol.~48, pp. 18--28, 1974.

\bibitem{cunche08:hybrid}
M.~Cunche and V.~Roca, ``Improving the decoding of ldpc codes for the packet
  erasure channel with a hybrid zyablov iterative decoding/gaussian elimination
  scheme,'' INRIA Research Report RR-6473, Tech. Rep., march 2008.

\bibitem{paolini08:CommLetters}
E.~Paolini, G.~Liva, B.~Matuz, and M.~Chiani, ``Generalized ira erasure
  correcting codes for hybrid iterative/maximum likelihood decoding,''
  \emph{Communications Letters, IEEE}, vol.~12, no.~6, pp. 450--452, June 2008.

\bibitem{USPatent6486803}
A.~Haken, M.~Luby, G.~Horn, J.~Byers, J.~Persch, and M.~Mitzenmacher, ``{On
  demand encoding with a window},'' US Patent 6486803, Nov. 2002.

\bibitem{Rich-Shok-Urba}
T.~J. Richardson, M.~A. Shokrollahi, and R.~L. Urbanke, ``Design of capacity
  approaching irregular low density parity check codes,'' \emph{IEEE Trans.
  Inform. Theory}, vol.~47, no.~2, pp. 619--637, 2001.

\bibitem{Kasai2003}
K.~Kasai, T.~Shibuya, and K.~Sakaniwa, ``Detailed representation of irregular
  ldpc code ensembles and density evolution,'' \emph{Information Theory, 2003.
  Proceedings. IEEE International Symposium on}, pp. 121--, June-4 July 2003.

\bibitem{Golub:1996}
G.~H. Golub and C.~F. van Loan, \emph{Matrix Computations}, 3rd~ed.\hskip 1em
  plus 0.5em minus 0.4em\relax Johns Hopkins University Press, 1996.

\bibitem{cunche08:gaussian}
M.~Cunche and V.~Roca, ``Optimizing the error recovery capabilities of
  ldpc-staircase codes featuring a gaussian elimination decoding scheme,'' in
  \emph{10th IEEE International Workshop on Signal Processing for Space
  Communications (SPSC'08)}, october 2008.

\end{thebibliography}
%
%
%
%

\end{document}